\documentclass{elsart}
 \usepackage{graphicx}
 \usepackage{epsfig}

\usepackage{amssymb}

\begin{document}

\begin{frontmatter}

\title{Universal Rise of Hadronic 
Total Cross Sections based on Forward $\pi p$ and $\bar pp(pp)$ Scatterings }

\author{Muneyuki Ishida$^a$ and Keiji Igi$^b$ }
\address{$^a$Department of Physics, School of Science and Engineering, Meisei University, Hino, Tokyo 191-8506, Japan}
\address{$^b$Theoretical Physics Laboratory, RIKEN, Wako, Saitama 351-0198, Japan}

\begin{abstract}
Recently there are several evidences of the increase of the total cross section 
$\sigma_{\rm tot}$ to be log$^2 s$ consistent with the Froissart unitarity bound, and the 
COMPETE collaborations in the PDG have further assumed  
$\sigma_{\rm tot}\simeq B$log$^2(s/s_0)$ to extend its universal rise with a common
value of $B$ for all the hadronic scatterings. 
However, there is no rigorous proof yet based only on QCD. Therefore, it is worthwhile 
to prove this universal rise of $\sigma_{\rm tot}$ even empirically. 
In this letter we attempt to obtain the value of $B$ for $\pi p$ scattering,
$B_{\pi p}$, with reasonable accuracy by taking into account 
the rich $\pi p$ data in all the energy regions. 
We use the finite-energy sum rule(FESR) expressed in terms of the $\pi p$ scattering 
data in the low and intermediate energies as a constraint between high-energy parameters.
We then have searched for the simultaneous best fit to the $\sigma_{\rm tot}$ and 
$\rho$ ratios, the ratios of the real to imaginary parts of the forward scattering amplitudes.
The lower energy data are included in the integral of FESR, 
the more precisely determined is the non-leading term 
such as log $s$, and then helps to determine the leading terms like log$^2 s$.   
We have derived the value of $B_{\pi p}$ as $B_{\pi p}=0.311\pm 0.044$mb. 
This value is to be compared with the value of $B$ for $\bar pp,pp$ scattering,
$B_{pp}$, in our previous analysis[11],
$B_{pp}=0.289\pm 0.023$mb.  Thus, our result appears to support the universality hypothesis.
\end{abstract}

\end{frontmatter}


\noindent \underline{\it Purpose of this Letter}

It is well known that the increase of the total cross sections $\sigma_{\rm tot}$ 
is at most log$^2 s$ as the Froissart unitarity bound\cite{[1],[2]}. 
Recently, there have been several evidences\cite{[3],[4],[5],[6],[7]} 
to support the increase of $\sigma_{\rm tot}$ to be log$^2 s$. 
The COMPETE collaborations\cite{[4],[7]} have further assumed 
 $\sigma_{\rm tot}\simeq B$ log$^2(s/s_0)$ to extend the universal rise 
of all the total cross sections. 
That is, they took a common value of $B$ to fit all the data of  
$\bar pp(pp)$, $\pi^\pm p$, $K^\pm p$, $\Sigma^- p$, $\gamma p$ and $\gamma \gamma$
scatterings and this resulted in reducing the number of adjustable parameters. 
The universality of the coefficient $B$ 
was expected in the paper\cite{[Jenk]}, and
other theoretical supports\cite{[Itakura],[Strike]}
based on the arguments describing deep inelastic scattering by
gluon saturation in hadron light-cone wavefunction 
(the Colour Glass Condensate\cite{[CGC]} of QCD) 
were given in recent years. But there has been no rigorous proof yet based only on QCD.

Therefore, it is worthwhile to prove this universal rise even empirically. 
In the near future, the $pp$ total cross section $\sigma_{\rm tot}^{pp}$ will be measured 
at the LHC energy ($\sqrt s =14$TeV ) in TOTEM experiment.
Therefore, the value of $B$ for $\bar pp,pp$ scattering, $B_{pp}$, will be determined 
with good accuracy. On the other hand, the $\pi p$  total cross sections 
$\sigma_{\rm tot}^{\pi p}$ have been measured only up to $k$=610GeV, where
$k$ is the laboratory momentum of $\pi$ and it corresponds to $\sqrt s$=33.8GeV, 
by the SELEX collaboration\cite{[9]}. Thus, one might doubt to obtain the value of
$B$ for $\pi p$ scattering, $B_{\pi p}$, with reasonable accuracy.
 
The purpose of this letter is to attack this problem and to compare the values of 
$B_{pp}$ and $B_{\pi p}$ in a new light. 
We can use the rich informations of the experimental $\sigma_{\rm tot}$ data
 in the low energy regions through the finite-energy sum rule (FESR). 
We adopt the FESR with the integral region between 
$k=\overline{N_1}$ and $\overline{N_2}$\cite{[13]} as a constraint between
high-energy parameters, and analyze the $\pi^\mp p$ total cross sections 
$\sigma_{\rm tot}^{\pi^\mp p}$ and $\rho$ ratios $\rho^{\pi^\mp p}$,
the ratios of real to imaginary parts of the forward scattering amplitudes.
This FESR requires that the low-energy extension of the high-energy asymptotic formula
should coincide, roughly speaking, with the average of experimental $\sigma_{\rm tot}$
in the relevant region between $k=\overline{N_1}$ and $\overline{N_2}$.
This is called FESR duality.  
We have already used\cite{[8]}
\footnote{In our previous work\cite{[8]},
we also used the FESR of $P^\prime$ type\cite{[10],[11],[12]} which
includes the integral of $\sigma_{\rm tot}$ from the $\pi p$ threshold,
that is, $\overline{N_1}=0$ GeV. However, this sum rule needs one subtraction,
and the subtraction term $-F^{(+)}(0)$ should have been added to the LHS of 
Eq.(9) in ref.\cite{[8]}. The sum rule (14) for $\pi^\pm p$ in the same reference 
should be slightly modified since $F^{(+)}(0)=0$ has been assumed implicitly.
In the present analysis we do not use this implicit assumption.  
}
 this sum rule between $\overline{N_1}$=10GeV and 
$\overline{N_2}$=20GeV. The rich data in $k<10$GeV were not included
in this case, however.
The lower energy data are included in the integral of $\sigma_{\rm tot}$, 
the more precisely determined is the sub-leading term, i.e., the $P^\prime$ term
(the term with coefficient $\beta_{P^\prime}$ in Eq.~(\ref{eq1})),
which is built in the sense of FESR\cite{[10],[11],[12]} by the sum of 
direct channel resonances. Then, it helps to determine the non-leading term 
such as log $s$  which then helps to determine the leading term like log$^2 s$.  
Thus, in the present work we extend maximally the energy region of the input data
to take $\overline{N_1}\le 10$GeV, so as to obtain the value of $B_{\pi p}$
as most accurately as possible.

The $\bar pp$ scattering has open (meson) channels in the so-called unphysical regions
 with $\sqrt s < 2M$ ($M$ being the proton mass), and it may cause some trouble
in applying the FESR. 
A possible solution for $\bar pp$ will be discussed later in Eq.~(\ref{eq5}).
In contrast, there are no such effects in $\pi p$ scattering. 
Thus, we can take into account more resonances through FESR in order to obtain 
the low-energy extension from the high-energy side with good accuracy. 
To obtain a sufficiently small error of $B_{\pi p}$, it appears to be important to 
include the information of the low-energy scattering data with $0\le k \le 10$GeV 
through FESR.

We will show that the resulting value of  $B_{\pi p}$ is consistent with that of $B_{pp}$, which appears to support the universality 
hypothesis.

\noindent \underline{\it Analysis of Forward $\pi^\mp p$ Scattering}

In the following, we use the laboratory energy of the incident pion, denoted as $\nu$,
instead of the center of mass energy squared, $s$. They are related through
\begin{eqnarray}
s &=& 2M\nu + M^2+\mu^2
\label{nu}
\end{eqnarray}
with each other where $M(\mu )$ is proton(pion) mass. By using the variable $\nu$,
a crossing transformation is expressed exactly by $\nu \rightarrow -\nu$
in forward scattering amplitudes.

We take both the crossing-even and crossing-odd forward scattering amplitudes,
$F^{(+)}(\nu )$ and $F^{(-)}(\nu )$, which are defined from forward $\pi^{\mp}p$ scattering
amplitudes $f^{\pi^\mp p}(\nu )$ by 
\begin{eqnarray}
 F^{(\pm )}(\nu ) &=& ({f^{\pi^- p}(\nu )\pm f^{\pi^+ p}(\nu )})/{2} \ .
\label{eqF}
\end{eqnarray}
We assume
\begin{eqnarray}
{\rm Im}F^{(+)}(\nu ) 
 & \simeq & \frac{\nu}{\mu^2}  \left( c_0  +  c_1 {\rm log}\frac{\nu }{\mu} 
     + c_2 {\rm log}^2\frac{\nu }{\mu}  \right)
    + \frac{\beta_{P^\prime}}{\mu} \left( \frac{\nu}{\mu}\right)^{\alpha_{P^\prime}}\ 
\label{eq1}\\
{\rm Im}F^{(-)}(\nu ) & \simeq &  
     \frac{\beta_V}{\mu} \left( \frac{\nu}{\mu}\right)^{\alpha_V}\ ,
\label{eq2}
\end{eqnarray}
which are expected to be valid in the asymptotically high-energy region
above some energy  $\nu > N$.
$\nu$ is related with $k$ by $\nu =\sqrt{k^2+\mu^2}$, and the momentum 
corresponding to $\nu =N$ is represented by the quantity with overline such as 
$k=\overline{N}$ in this letter.
The imaginary parts are related to the total cross sections $\sigma_{\rm tot}^{(\pm )}$ by
the formula Im $F^{(\pm)}(\nu )=\frac{k}{4\pi}\sigma_{\rm tot}^{(\pm)}$,
and $\sigma_{\rm tot}^{\pi^\mp p}$ is given by 
$\sigma_{\rm tot}^{\pi^\mp p} =\sigma_{\rm tot}^{(+)}\pm \sigma_{\rm tot}^{(-)}$. 
These formulas (\ref{eq1}) and (\ref{eq2}) are derived 
by traditional Pomeron-Reggeon exchange model 
except for the terms with coefficients $c_2$ and $c_1$.
The coupling coefficients $\beta_{P^\prime}, c_0, \beta_V$ are 
the unknown parameters in the Regge theory. 
%
The $\alpha_{P^\prime},\alpha_V$ are determined phenomenologically by the
intercepts of Regge trajectories of $f_2(1275),\rho (770)$.
The $c_2,c_1$ terms are introduced consistently with Froissart bound to describe 
the rise of $\sigma_{\rm tot}$ in high-energy regions.   

By using the crossing property $F^{(\pm )}(-\nu ) = \pm F^{(\pm )}(\nu)^*$, 
the real parts are given by
\begin{eqnarray}
{\rm Re}F^{(+)}(\nu ) & \simeq & \frac{\pi \nu}{2\mu^2}\left( 
c_1 + 2 c_2 {\rm ln}\frac{\nu}{\mu} \right)
  -\frac{\beta_{P^\prime}}{\mu}\left(\frac{\nu}{\mu}\right)^{\alpha_{P^\prime}}
  {\rm cot}\frac{\pi\alpha_{P^\prime}}{2} + F^{(+)}(0)\ , \label{eq3}\\
{\rm Re}F^{(-)}(\nu ) & \simeq & \frac{\beta_V}{\mu}\left(\frac{\nu}{\mu}\right)^{\alpha_V}
  {\rm tan}\frac{\pi\alpha_V}{2} \ ,
\label{eq4}
\end{eqnarray}
where $F^{(+)}(0)$ is a subtraction constant. The equations (\ref{eq3}) and (\ref{eq4})
are used in fit to $\rho$ ratios, 
$\rho^{\pi^\mp p}$ $=$
 ${\rm Re}\ f^{\pi^\mp p}/{\rm Im}\ f^{\pi^\mp p}$.

We fit the experimental $\sigma_{\rm tot}^{\pi^\mp p}$
and $\rho^{\pi^\mp p}$ ratios simultaneously. 
The $c_{2,1,0}$, $\beta_{P^\prime ,V}$ and $F^{(+)}(0)$ are parameters, while
the $\alpha_{P^\prime}\simeq \alpha_V$ is taken to be the empirical value $\simeq 0.5$.
The $B_{\pi p}$ is related with the 
dimensionless parameter $c_2$ by $B_{\pi p}=\frac{4\pi}{\mu^2}c_2$ in unit of mb.

The FESR is used as a constraint between these parameters\cite{[13],[8]}.
\begin{eqnarray}
\frac{2}{\pi}\int_{N_1}^{N_2} && \frac{\nu }{k^2} 
 \ {\rm Im}F^{(+)}(\nu )\ d\nu 
  = \frac{1}{2\pi^2} \int_{\overline{N}_1}^{\overline{N}_2} 
\ \sigma^{(+)}_{\rm tot}(k )\ dk \ \ ,\ \ \ \ \ \ \ 
\label{eq5}
\end{eqnarray}
where the laboratory energies ${N_{1,2}}$ are related to the corresponding momenta
$\overline{N_{1,2}}$ by ${N_{1,2}}=\sqrt{\overline{N_{1,2}}^2+\mu^2}$
as explained above. The value of $\overline{N_2}$ should be selected to be reasonably
high momentum above which no resonance structures are observed, 
while $\overline{N_1}$ may be taken to be in the resonance energy
region in the sense of FESR duality. 
%

The integrand of the LHS of Eq.~(\ref{eq5}) is the low-energy extension of
Eqs.~(\ref{eq1}).
The RHS is the integral of experimental 
$\sigma_{\rm tot}^{(+)}(=(\sigma_{\rm tot}^{\pi^- p}+\sigma_{\rm tot}^{\pi^+p})/2)$ 
in the resonance energy regions.
This shows up several peak and dip structures corresponding to a number of $N$ and 
$\Delta$ resonances, in addition to the non-resonating background. 
Thus, Eq.~(\ref{eq5}) means the FESR duality, that is, 
the average of these resonance structures plus the non-resonating background 
in $\sigma_{\rm tot}^{(+)}$ should coincide with the low-energy extension of 
the asymptotic formula.
%
Practically, the RHS can be estimated from the experimental 
$\sigma_{\rm tot}^{\pi^\mp p}$ very accurately with errors less than 0.5\% ,
so we can use Eq.~(\ref{eq5}) as an exact constraint.

In case of $\bar pp,pp$ scattering, 
if we take too small value of $N_1$ close to the threshold $\nu =M$,
the FESR (\ref{eq5}) is affected strongly by a contribution 
from the unphysical region $\nu <M$, and often does not work well. 
Thus, we must take $N_1$ to be fairly larger than $M$.
In contrast, there is no such problem in $\pi p$ scattering.
The lower the value of $\overline{N_1}$ is taken, the more the information of low energy
scattering data are included. Then, the more accurately estimated value of $c_2$ is obtained. 
We try to take $\overline{N_1}$ as small value as possible in the present analysis. 
\\

\begin{table*}
\caption{
Values of parameters in the best fit with five-parameters, using FESR as a constraint, 
where the value of $\beta_{P^\prime}$ is obtained from FESR constraint and
$(\alpha_{P^\prime},\alpha_V)$ is fixed to be $(0.500,0.497)$.
The statistical errors of $c_2$ are also given.
The result of six-parameter fit 
without using FESR is also shown in the last row as No SR.
}
\begin{tabular}{l|cccccc}
$\overline{N_1}$(GeV) &  $c_2\times 10^5$ & $c_1$  & $c_0$ & $F^{(+)}(0)$ & $\beta_V$ & $\beta_{P^\prime}$\\
\hline
10 & 126$\pm$30
 & -0.0125 & 0.117 & -0.321 & 0.0389 &  0.136 \\
7 & 128$\pm$26
 & -0.0128 & 0.118 & -0.384 & 0.0389 &  0.132 \\
5 & 127$\pm$24
 & -0.0128 & 0.118 & -0.333 & 0.0388 &  0.133 \\
4 & 126$\pm$22
 & -0.0125 & 0.117 & -0.239 & 0.0388 &  0.137 \\
3.02 & 123$\pm$21
 & -0.0120 & 0.115 & -0.043 & 0.0388 &  0.126 \\
2.035 & 119$\pm$20
 & -0.0112 & 0.111 & 0.252 & 0.0388 &  0.137 \\
1.476 & 118$\pm$19
 & -0.0111 & 0.110 & 0.285 & 0.0388 &  0.139 \\
0.9958 & 119$\pm$18
 & -0.0112 & 0.111 & 0.247 & 0.0388 &  0.137 \\
0.818 & 124$\pm$18
 & -0.0122 & 0.115 & -0.069 & 0.0388 &  0.125 \\
0.723 & 129$\pm$17
 & -0.0131 & 0.120 & -0.347 & 0.0388 &  0.114 \\
0.475 & 143$\pm$17
 & -0.0155 & 0.131 & -1.111 & 0.0387 &  0.084 \\
0.281 & 126$\pm$16
 & -0.0124 & 0.116 & -0.123 & 0.0388 &  0.122 \\
\hline
No SR & 95$\pm$45
 & -0.0069 & 0.091 & 1.643 & 0.0390 &  0.209 \\
\hline
\end{tabular}
\label{tab1}
\end{table*}

\begin{table*}
\caption{
Values of the best-fit $\chi^2$ for each case.
The FESR is used as a constraint, and 
five-parameter($N_P$=5) fit is performed.
Both total $\chi^2$ and respective $\chi^2$ for each data with the number of data points 
are given. The $\chi^2$ of six-parameter($N_P$=6) fit 
without using FESR is also shown in the last row as No SR.
}
\begin{tabular}{l|c|cccc}
$\overline{N_1}$(GeV) &  $\frac{\chi^2_{\rm tot}}{N_D-N_P}$
     & $\frac{\chi^{2,\sigma}_{\pi^-p}}{N^\sigma_{\pi^-p}}$
     & $\frac{\chi^{2,\rho}_{\pi^-p}}{N^\rho_{\pi^-p}}$
     & $\frac{\chi^{2,\sigma}_{\pi^+ p}}{N^\sigma_{\pi^+ p}}$
     & $\frac{\chi^{2,\rho}_{\pi^+ p}}{N^\rho_{\pi^+ p}}$ \\
\hline
10 & $\frac{72.58}{162-5}$ & $\frac{12.47}{84}$ & $\frac{40.77}{33}$ & $\frac{6.66}{37}$ & $\frac{12.68}{8}$\\
7 & $\frac{72.56}{162-5}$ & $\frac{12.44}{84}$ & $\frac{40.76}{33}$ & $\frac{6.69}{37}$ & $\frac{12.67}{8}$\\
5 & $\frac{72.49}{162-5}$ & $\frac{12.53}{84}$ & $\frac{40.65}{33}$ & $\frac{6.67}{37}$ & $\frac{12.64}{8}$\\
4 & $\frac{72.42}{162-5}$ & $\frac{12.64}{84}$ & $\frac{40.51}{33}$ & $\frac{6.64}{37}$ & $\frac{12.63}{8}$\\
3.02 & $\frac{72.29}{162-5}$ & $\frac{12.87}{84}$ & $\frac{40.26}{33}$ & $\frac{6.56}{37}$ & $\frac{12.60}{8}$\\
2.035 & $\frac{72.12}{162-5}$ & $\frac{13.22}{84}$ & $\frac{39.89}{33}$ & $\frac{6.46}{37}$ & $\frac{12.55}{8}$\\
1.476 & $\frac{72.10}{162-5}$ & $\frac{13.27}{84}$ & $\frac{39.85}{33}$ & $\frac{6.45}{37}$ & $\frac{12.55}{8}$\\
0.9958 & $\frac{72.12}{162-5}$ & $\frac{13.24}{84}$ & $\frac{39.88}{33}$ & $\frac{6.46}{37}$ & $\frac{12.54}{8}$\\
0.818 & $\frac{72.29}{162-5}$ & $\frac{12.91}{84}$ & $\frac{40.24}{33}$ & $\frac{6.56}{37}$ & $\frac{12.57}{8}$\\
0.723 & $\frac{72.46}{162-5}$ & $\frac{12.64}{84}$ & $\frac{40.57}{33}$ & $\frac{6.66}{37}$ & $\frac{12.60}{8}$\\
0.475 & $\frac{73.08}{162-5}$ & $\frac{11.96}{84}$ & $\frac{41.48}{33}$ & $\frac{6.98}{37}$ & $\frac{12.66}{8}$\\
0.281 & $\frac{72.32}{162-5}$ & $\frac{12.89}{84}$ & $\frac{40.29}{33}$ & $\frac{6.58}{37}$ & $\frac{12.56}{8}$\\
\hline
No SR & $\frac{71.79}{162-6}$ & $\frac{14.94}{84}$ & $\frac{38.35}{33}$ & $\frac{6.09}{37}$ & $\frac{12.41}{8}$\\
\hline
\end{tabular}
\label{tab2}
\end{table*}

\noindent \underline{\it Result of the analyses}

The data\cite{[7]} of $\sigma_{\rm tot}^{\pi^\mp p}$ for $k\ge 20$ GeV and $\rho^{\pi^\mp p}$ for $k\ge 5$ GeV
are fitted simultaneously.
In the FESR, Eq.~(\ref{eq5}), $\overline{N_2}$ is taken to be 20GeV.
The values of $\overline{N_1}$ are taken to be
10, 7, 5, 4, 3.02, 2.035, 1.476, 0.9958, 0.818, 0.723, 0.475, 0.281 GeV.
Except for the first three values, they correspond to the energies of peak and dip positions 
of experimental $\sigma_{\rm tot}^{\pi^- p}$ or $\sigma_{\rm tot}^{\pi^+ p}$.
For each value of $\overline{N_1}$, the FESR is derived. It is used as a constraint 
between the parameters, $c_{2,1,0}$ and $\beta_{P^\prime}$, and the fitting is 
performed. The number of fitting parameters is five, including $\beta_V$ and $F^{(+)}(0)$.
The $(\alpha_{P^\prime},\alpha_V)$ are fixed to be 
$(0.500,0.497)$\cite{[8]} in all the fitting procedures.
The values of parameters and $\chi^2$ in the best fits in respective cases are 
given in Tables \ref{tab1} and \ref{tab2}.

It is remarkable that the values of the parameters in the best fits
are almost independent of $\overline{N_1}$ (except for the case of 0.475GeV), 
as can be seen in Table \ref{tab1}. 
The results are surprisingly stable, although  
there are many resonant strucures observed and $\sigma_{\rm tot}$ 
show sharp peak and dip structures in this energy region. 
The lower the value of $\overline{N_1}$ is taken, 
the smaller the statistical errors of $c_2$ become in the best fits.
We can adopt the case of   
$\overline{N_1}$=0.818GeV as the representative of our results.
The value of $c_2$ in the best fit is
\begin{eqnarray}
c_2 &=& (124\pm 18)\cdot 10^{-5}\ \ .
\label{eq6}
\end{eqnarray}
The central value of Eq.~(\ref{eq6}) is almost the same as (126$\pm$30) 
in the case of $\overline{N_1}$=10GeV, but the error is 
much improved.\footnote{
The $c_2\ ($log~$\nu )^2 + c_1$~log~$\nu$ with $c_2>0$ shows the shape of parabola 
as a function of log~$\nu$ with minimum. Therefore, the information on the low-energy 
side is helpful in order to obtain the accuracy of the value of $c_2$. 
} 
This shows that the data with $k\le 10$GeV give very important information
to determine the high-energy parameters such as $c_2$ through the FESR duality. 

\noindent \underline{\it Concluding Remarks}

Using the value of $c_2$ in Eq.~(\ref{eq6}), we can derive the value of $B_{\pi p}$ as 
\begin{eqnarray}
B_{\pi p} &=& \frac{4\pi}{\mu^2}c_2=0.311\pm 0.044{\rm mb}\ \ .
\label{eq7}
\end{eqnarray}
This value is to be compared with the value of $B_{pp}$ in our previous analysis,
$B_{pp}=0.289\pm 0.023$mb\cite{[8]}. 
The $B_{\pi p}$ in Eq.~(\ref{eq7}) is consistent with this $B_{pp}$. 
Thus, our result appears to support the universality hypothesis for the 
values of $B$ parameters. 

In case of six parameter fit without using FESR, we obtain 
$c_2$=(95$\pm$45)$\cdot 10^{-5}$, shown in the last row of 
Table \ref{tab1}. This value corresponds to $B_{\pi p}=0.24\pm 0.11$mb.
From this value, we would not be able to say anything about 
the universality due to its large statistical error.  
The role of FESR is crucially important to obtain a definite conclusion.

It is to be noted that our value of $B_{\pi p}$ is consistent with 
the value of $B$ by COMPETE collab.\cite{[4],[7]}, $0.308\pm 0.010$mb, which is obtained 
by assuming the universality of $B$ for various processes.

\end{document}